# A series circuit of thermal rectifiers: an effective way to enhance rectification ratio

Shiqian Hu[1], Meng An[2,3], Nuo Yang[2,3*], and Baowen Li[4*]

[1]Center for Phononics and Thermal Energy Science, School of Physics Science and Engineering, Tongji University, Shanghai 200092, People's Republic of China

[2]State Key Laboratory of Coal Combustion, Huazhong University of Science and Technology (HUST), Wuhan 430074, People's Republic of China

[3]Nano Interface Center for Energy (NICE), School of Energy and Power Engineering, Huazhong University of Science and Technology (HUST), Wuhan 430074, People's Republic of China

[4]Department of Mechanical Engineering, University of Colorado, Boulder, CO 80309

*Corresponding authors:
Email: nuo@hust.edu.cn (N.Y.)

Email: Baowen.Li@Colorado.Edu (B.L.)

**Abstract**

The low rectification ratio limits the application of thermal rectifiers. In this paper, it is found that a series circuit of thermal rectifiers in asymmetric graphene/graphene phononic crystal (GPnC) structures can improve rectification significantly. The simulations by using non-equilibrium molecular dynamics (NEMD) are performed. Moreover, the size effect on thermal rectification is also studied and discussed.

**Introduction**

Phonon, the quantized mode of vibration, is responsible for heat transport in dielectric and semiconducting crystals. The understanding of phonon properties provides a theoretical foundation for thermal devices, such as thermal rectifiers,[1-9] optomechanical crystals,[10,11] thermal cloaking,[12-16] thermoelectrics,[17-21] thermocrystals[22-24] and phonon nanocapacitor[25] etc., which could lead to a new technological revolution.

The graphene, a two-dimensional structure of a single atomic layer of carbon arranged in a honeycomb lattice,[26] exhibits many intriguing properties. In addition to its unique electronic and optical properties, graphene is a promising material for thermal management of future nanoelectronic devices due to its superior thermal conductivity at room temperature.[5,27-29] In addition, the porous graphene have shown potential applications in many fields, such as gas separation and purification,[30] water desalination,[31] ion channel[32] and DNA sequencing.[33]

Similar to electronic counterpart, the thermal rectifier also plays a vital role in phononics/thermal circuits.[1] The thermal rectification means that the magnitude of heat flux changes when the temperature gradient is reversed in direction. Massive efforts have been devoted to investigate a high-performance thermal rectifier. Up to now, most theoretical studies and experimental tests are realized based on nonlinear lattice models,[6] graded mass density,[2,7] asymmetric structure[3-5,9,34] or an interface between two different materials.[35,36] Very recently, based on phase change materials,[37,38] the thermal rectification can be enhanced significantly. However, available reports to enhance the rectification ratio only focus on optimizing the performance of single thermal rectifier.

In this letter, a series circuit of thermal rectifier is proposed to enhance thermal rectification. Based on our previous work, the different temperature dependence of thermal conductivity in graphene and graphene phononic crystal (GPnC),[39] we construct an asymmetric graphene/GPnC structure (Fig. 1(a)) and found its thermal rectification. Furthermore, inspired by the series effect in electronic circuits, the graphene/GPnC$_1$/GPnC$_2$ structure (Fig. 2(c)) is used to investigate the series effect of thermal rectification. The dependence of the thermal rectification on the geometric size of structure is also discussed in details.

**Simulations and calculations**

A schematic picture of the asymmetric graphene/GPnC structure is shown in Fig. 1(a), which is comprised of graphene and GPnC with the same length. The size of pores in GPnC is characterized by $L_1$. The neck width ($L_n$) is fixed as 0.71 nm in all structures studied here. In our simulations, the lattice constant and thickness of graphene are used as 0.14 nm and 0.33 nm, respectively. The fixed (periodic) boundary condition is adopted along longitudinal (transversal) direction.

In non-equilibrium molecular dynamics (NEMD) simulations, the three layers atoms close to the two ends are coupled with the Langevin heat bath[40] at temperature $T_L$ =

$T_0(1+\Delta)$ and $T_R = T_0(1-\Delta)$, respectively, where $T_0$ is the average temperature, and $\Delta$ is the normalized temperature difference. Therefore, the atoms in the left end are at a higher (lower) temperature in the case of $\Delta > 0$ ($\Delta < 0$).

The bonding interaction between carbon atoms is described by the energy potential of a Morse bond and a harmonic cosine angle, which include both two-body and three-body potential terms.[41] We use the velocity Verlet algorithm to integrate the discretized differential equations of motions, where the time step, $\Delta t$, is set as 0.5 fs. The heat flux transferred across the system is defined as the energy transported along it in unit time.[3] The structure is relaxed long enough (3 ns) such that the system reaches the steady state. Then, both the temperature and heat flux are recorded, which are averaged over 4.5 ns. It is noted that the values of temperature difference ($\Delta T$) are consistent when the heat flux ($J_+/J_-$) is recorded. All the results presented here are averaged from 12 independent simulations from different initial conditions. The error bar is the standard errors of the results of these simulations.

To show the thermal rectification effect quantitatively, the thermal rectification ratio is defined as,

$$R_{ect} = \frac{r_-}{r_+}. \qquad (1)$$

where $r_+$ ($r_-$) is the thermal resistance of the system when $\Delta < 0$ ($>0$).

We define the generalized heat flux transmission function ($\tau_r$) as,[42]

$$\tau_{r\pm} = \frac{J_{\pm}}{k_B \Delta T}, \qquad (2)$$

where $J_+$ ($J_-$) is the heat current from the right to the left (the left to the right), corresponding to $\Delta < 0$ ($>0$), $k_B$ is the boltzmann constant and $\Delta T$ is temperature difference. In the ballistic limit, the transmission function is simply equal to 3N (the number of propagating phonon modes). And according the definition of thermal resistance ($r = \Delta T/J$), we can easily obtain the thermal rectification as,

$$R_{ect} = \frac{r_-}{r_+} = \frac{\tau_{r+}}{\tau_{r-}}. \qquad (3)$$

For the series of two thermal rectifiers, $TR_1$ and $TR_2$, it is assumed that they are independent and not interact with each other. And then we arrive at the constitutive thermal rectification ratio of the series rectifiers,

$$R_{ect}^{se} = \frac{r_{12-}}{r_{12+}} = \frac{\tau_{r1,+} \tau_{r2,+}}{\tau_{r1,-} \tau_{r2,-}} = R_{ect}^1 R_{ect}^2, \qquad (4)$$

where $R_{ect}^1$ ($R_{ect}^2$) are the rectification ratio of $TR_1$ ($TR_2$).

**Results and Discussions**

Firstly, the thermal rectification effect is studied in an asymmetric graphene/GPnC structure (Fig. 1(a)). The temperature profiles in the direction from GPnC to graphene and the inverse direction in asymmetric graphene/GPnC structure are both plotted in Fig. 1(b). The thermal rectification ratio (R) is shown in Fig. 1(c) as a function of average temperature ($T_0$). It is found that the heat flux prefers from GPnC to graphene and the thermal rectification ratio (R) reaches an optimum value when $T_0$ equals 1500

K. Therefore, we choose $T_0$ as 1500 K in the following studies. The dependence of R on the normalized temperature difference ($\Delta T$) at 1500 K is shown in Fig. 1(d). It is observed that the heat flux prefers from GPnC to graphene, and that the value of R increases when $\Delta$ increases from 0.1 to 0.9.

In the following, we give a discussion on the mechanism of rectification in asymmetric graphene/GPnC structure, based on the difference of thermal resistance. Compared with the graphene, the thermal resistance of GPnC has a less sensible dependence on temperature.[39] The difference between the total thermal resistances of the two directions originates mainly from the resistance of graphene region. When the heat transfers from GPnC to graphene, the graphene region has a lower temperature (~800 K, Fig. 1(b)), corresponding to a lower thermal resistance. For the inverse direction, from GPnC to graphene, the graphene region has a higher temperature (~2000 K, Fig. 1(b)), corresponding to a higher thermal resistance. The smaller thermal resistance from GPnC to graphene means the heat flux preferring from GPnC to graphene.

Further, inspired by the series effect in electronic rectifiers, the basic element can be composed into complex structure to enhance the rectification ratio. We proposed a three section structure, graphene/GPnC$_1$/GPnC$_2$ (shown in Fig. 2(c)), where the size of pores in the GPnC$_1$ (GPnC$_2$) region is characterized by $L_1$ ($L_2$). The GPnC$_1$ and GPnC$_2$ have the same length. In the three section structure, the length of the graphene and GPnC are set as $L_{graphene}$ and $L_{GPnC}$ ($L_{GPnC_1} + L_{GPnC_2}$) respectively.

The results of thermal rectification of graphene/$GPnC_1$/$GPnC_2$ versus $\Delta$ are shown in Fig. 3 (black squares). Similar to asymmetric graphene/GPnC, the heat flux also prefers from GPnC to graphene and the value of thermal rectification ratio increases when $\Delta$ increases from 0.1 to 0.9. Meanwhile, the interesting finding is that the thermal rectification ratio of graphene/$GPnC_1$/$GPnC_2$ is enhanced significantly compared with that of asymmetric graphene/GPnC. For a comparison with graphene/$GPnC_1$/$GPnC_2$, the structure of graphene/$GPnC_2$/$GPnC_1$ is also studied. The results are also shown in Fig. 3 (blue triangles). Instead of enhancement, the value of thermal rectification ratio is reduced.

To interpret this enhancement (reduction) of R by the series effect in three section structures, we studied the thermal rectification of the graphene/$GPnC_1$ (Fig. 2(a)) and the $GPnC_1$/$GPnC_2$ (Fig. 2(b)). For the $GPnC_1$/$GPnC_2$ structure, the results show that the heat flux prefers from $GPnC_2$ to $GPnC_1$ (not shown here). When three section structures is arranged as graphene/$GPnC_1$/$GPnC_2$ (graphene/$GPnC_2$/$GPnC_1$), these two TRs are linked on the same (inverse) thermal rectification direction, which is responsible for the thermal rectification enhancement (reduction).

There are two sets of results for three section structures. The first one from series effect prediction by Eq. (4) (pink circles in Fig. 3) based on MD simulation of graphene/$GPnC_1$ and $GPnC_1$/$GPnC_2$. And the other one from MD simulation of three section structures (black squares and blue triangles in Fig. 3). The two sets results are agree well with each other as shown in Fig. 3. It demonstrates that, through a subtle arrangement of GPnCs, the thermal rectification ratio can be enhanced with series effect strategy.

Besides the series effect, the size effect should be another important factor of nanoscale devices. The ratio of $L_{graphene}/L_{GPnC}$ is defined as $P_L$. Then, the effect of length ratio $P_L$ in asymmetric graphene/GPnC on the rectification is investigated. When $P_L$ is changed from 1/3 to 3, it is found that the thermal rectification ratio increase with $P_L$ increase (shown in Fig. 4(a)). As mentioned in the above, the difference of the total thermal resistances come mainly from the resistance of graphene region. For graphene, the thermal conductivity ($\kappa_G$) increases with its length (L), as $\kappa_G \sim \log L$, which has been shown in recent experiment study.[43] As the length ratio ($P_L$) increases, due to size effect, the graphene region will have a larger thermal conductivity and smaller resistance. As shown in Fig. 4(b), the heat flux increase as $P_L$ increases, which results in a larger difference of resistance and thermal rectification ratio.

Considering both the series effect and size effect, we show in Fig. 5 that the thermal rectification of optimized graphene/$GPnC_1$/$GPnC_2$ structures, whose parameters are $P_L$ = 3 and $L_1$ = 0.71 nm and $L_2$ = 1.55 nm or 2.41 nm, respectively.

**Summary**

In summary, using non-equilibrium molecular dynamics simulations, we studied thermal rectification in both two section and three section asymmetric graphene/GPnC structures (Fig. 2). The series effect is demonstrated by the consistence between the results of theoretical prediction and that of MD simulations (Fig. 3). It is found that both the series effect and size effect are effective approaches to enhance the thermal rectification ratio, which can enhance the thermal rectification ratio by around 4 times (Fig. 5). The series model advances previous thermal rectifier designs. More recently, the large scale synthesis[44] of porous graphene sheet with pores size ranging from atomic precision (<1 nm) to nano-scaled dimension (~1-500 nm) has been demonstrated experimentally, therefore, the high-performance thermal rectifier proposed in this paper could be realized in experimentally in a foreseeable future.

**Acknowledgements**

This project was supported in part by the grants from the National Natural Science Foundation of China: 11334007 and 51576076. The authors thank the National Supercomputing Center in Tianjin (NSCC-TJ) for providing assistance in computations.

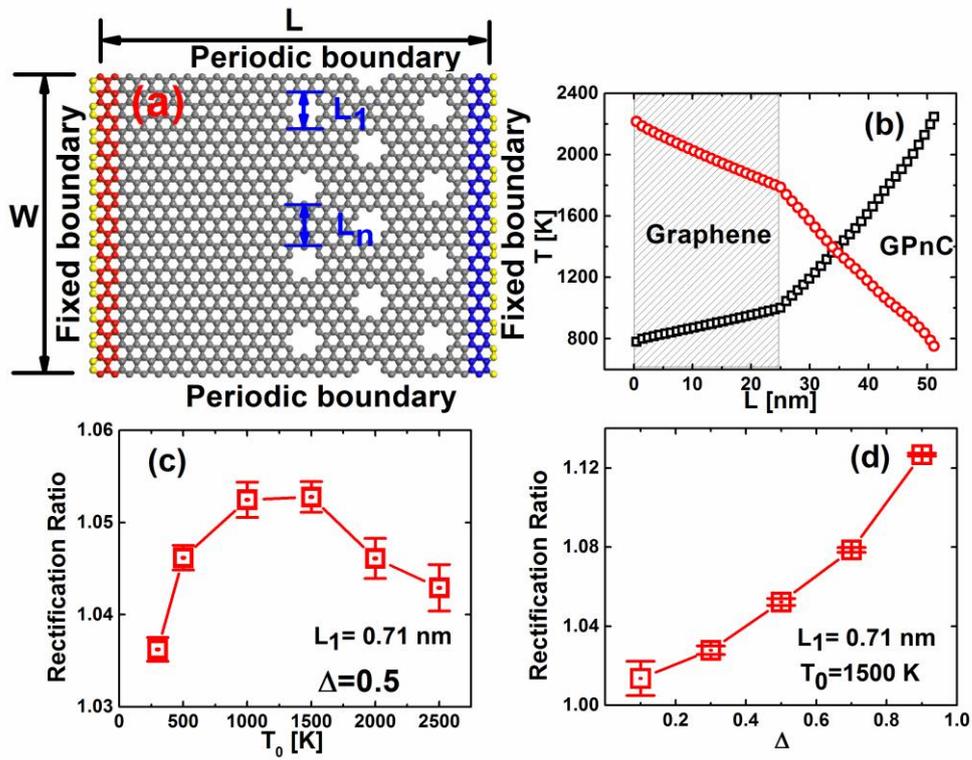

Fig. 1. (a) Schematic picture of asymmetric graphene/GPnC structure. The length (width) of the GPnC is set as L (W). $L_1$ (size of pores) and $L_n$ (neck width) are used to characterize the GPnC structure. $L_n$ is fixed as 0.71 nm. (b) The temperature profiles correspond to the heat flux along the direction from GPnC to graphene and the inverse direction in the asymmetric graphene/GPnC structure. (c) The thermal rectification ratio as a function of the average temperature $T_0$ with $\triangle = 0.5$. (d) The dependence of thermal rectification ratio on $\triangle$. The corresponding parameters are L = 50 nm, W = 7 nm, $L_1$ = 0.71 nm, $T_0$ = 1500 K.

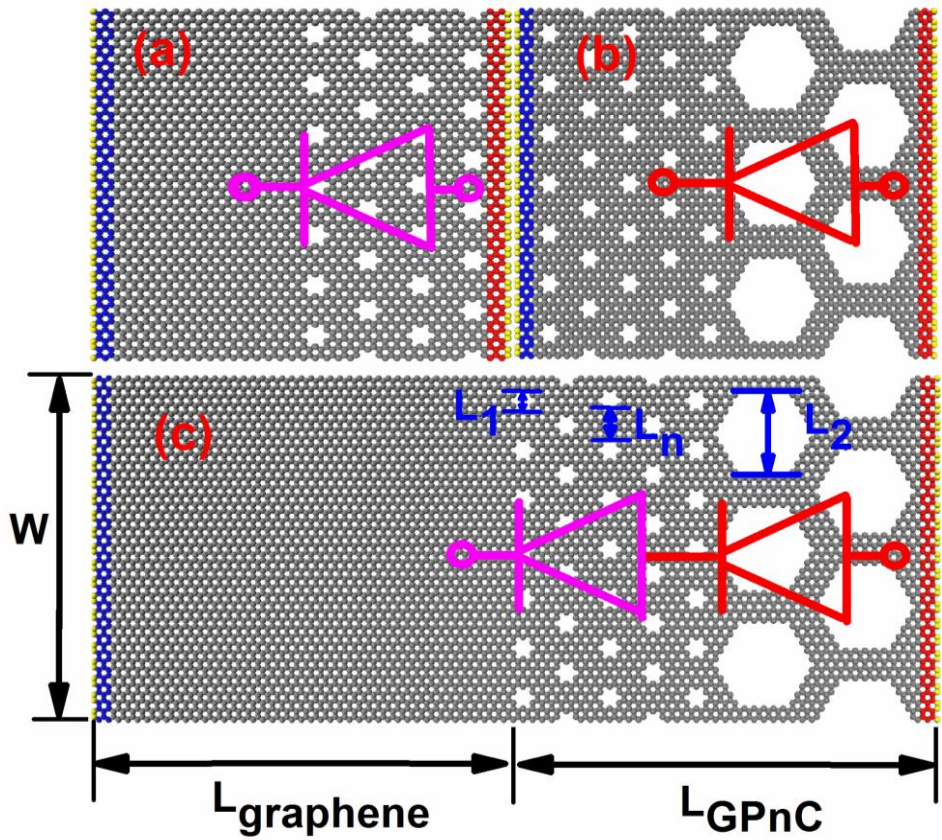

Fig. 2. (Color on-line) Schematic pictures of asymmetric graphene/GPnC structure (a), GPnC$_1$/GPnC$_2$ structure (b), and graphene/GPnC$_1$/GPnC$_2$ structure (c). L = L$_{graphene}$ + L$_{GPnC}$, and P$_L$ = L$_{graphene}$/L$_{GPnC}$. For all the three structures, the widths (W) are set as 7 nm. The fixed (periodic) boundary condition is applied along the transverse (longitudinal) direction. The sizes of pores in the middle (right) region are characterized by L$_1$ (L$_2$).

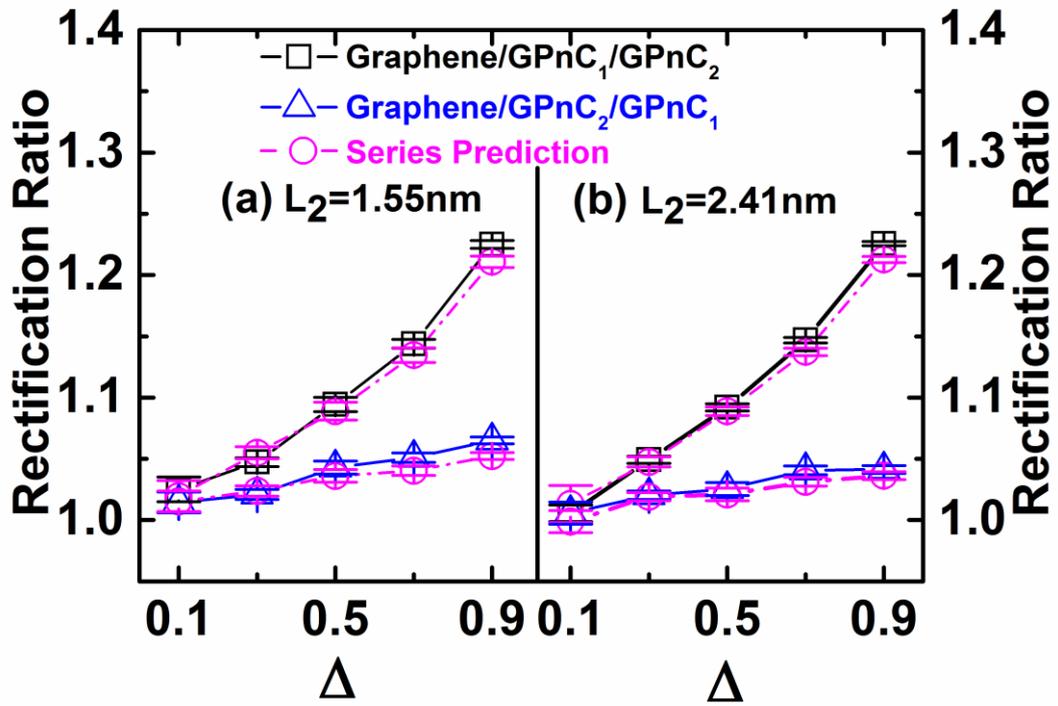

Fig. 3. (Color on-line) Rectification ratio versus $\Delta$ in the graphene/GPnC$_1$/GPnC$_2$ (graphene/GPnC$_2$/GPnC$_1$) structure corresponding to black squares (blue triangles). Pink circles are predicted results based on series effects Eq. (3). The $\triangle$ increases from 0.1 to 0.9. The parameters are L = 50 nm, W = 7 nm, L$_n$ = 0.71 nm, P$_L$ = 1, T$_0$ = 1500 K, L$_1$ = 0.71 nm, L$_2$ are 1.55 nm and 2.41 nm corresponding to (a) and (b) respectively. The error bar is standard errors of 12 MD simulations with different initial conditions.

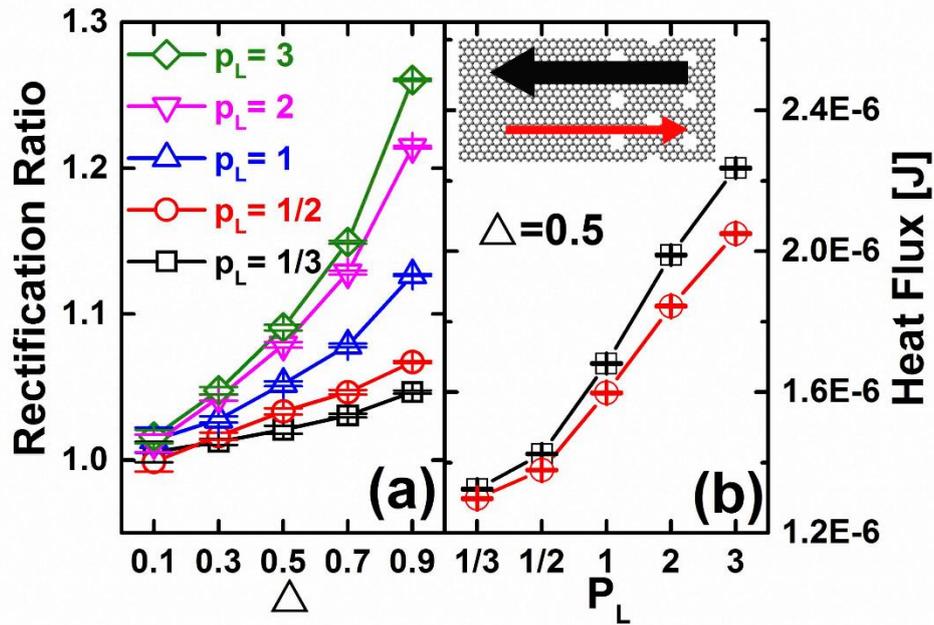

Fig. 4. (Color on-line) (a) Rectification ratio versus $\Delta$ in the asymmetric graphene/GPnC structures. The value of thermal rectification ratio increases as the increase of $P_L$. (b) Heat flux along the direction from GPnC to graphene and the inverse direction in asymmetric graphene/GPnC structure versus the length ratio ($P_L$). It shows that heat flux (thermal resistance) increase (decrease) with $P_L$ increase. The $\Delta$ increases from 0.1 to 0.9. $P_L$ is in the range from 1/3 to 3. The structure parameters are $L = 50$ nm, $W = 7$ nm, $L_n = 0.71$ nm, $L_l = 0.71$ nm, and $T_0 = 1500$ K. The error bar is standard errors of 12 MD simulations with different initial conditions.

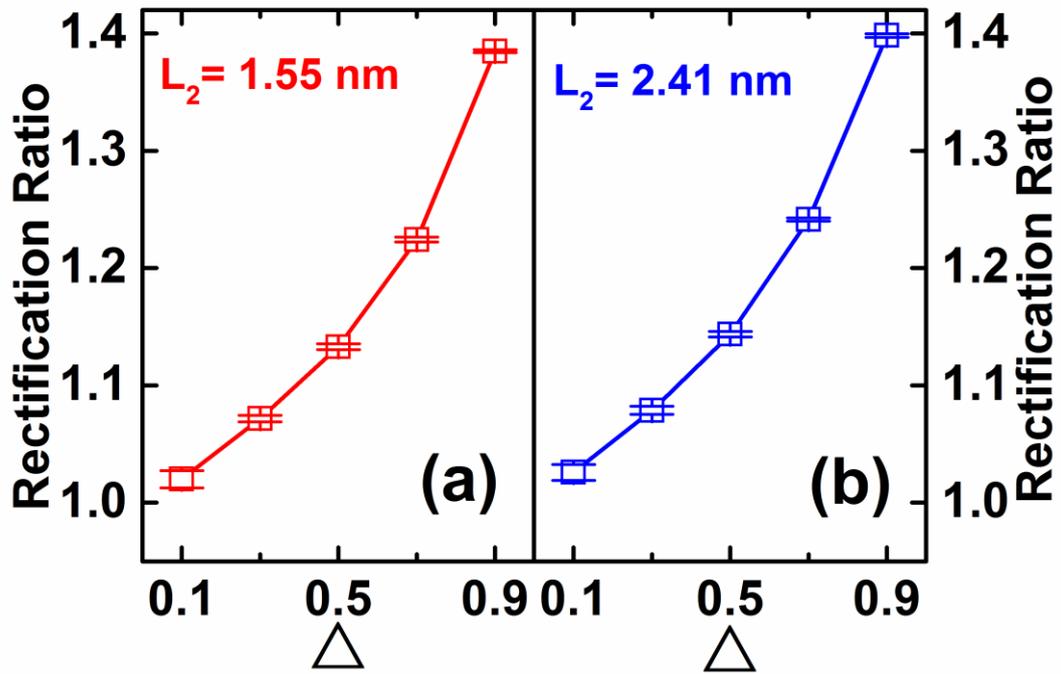

Fig. 5. (Color on-line) Rectification ratio versus $\Delta$ in the graphene/GPnC$_1$/GPnC$_2$ structure. The $\triangle$ increases from 0.1 to 0.9. The structure parameters are L = 60 nm, W = 7 nm, L$_n$ = 0.71 nm, P$_L$ = 3, T$_0$ = 1500 K, L$_1$ = 0.71 nm, L$_2$ are 1.55 nm and 2.41 nm corresponding to (a) and (b) respectively. The error bar is standard errors of 12 MD simulations with different initial conditions.